\documentclass[twocolumn, switch]{article}
\usepackage{preprint}

\usepackage{amsmath,amssymb,amsfonts}
\usepackage[numbers,square]{natbib}
\usepackage[utf8]{inputenc}
\usepackage[T1]{fontenc}
\usepackage{xcolor}
\usepackage{float}
\usepackage[colorlinks = true,
            linkcolor = purple,
            urlcolor  = blue,
            citecolor = cyan,
            anchorcolor = black]{hyperref}
\usepackage{authblk}

\usepackage{titlesec}
\titlespacing\section{0pt}{12pt plus 3pt minus 3pt}{1pt plus 1pt minus 1pt}
\titlespacing\subsection{0pt}{10pt plus 3pt minus 3pt}{1pt plus 1pt minus 1pt}
\titlespacing\subsubsection{0pt}{8pt plus 3pt minus 3pt}{1pt plus 1pt minus 1pt}

\title{Ultrafast Dipolar Electrostatic Modeling of Plasmonic Nanoparticles with Arbitrary Geometry}

\usepackage{tikz,xcolor}

\definecolor{lime}{HTML}{A6CE39}
\DeclareRobustCommand{\orcidicon}{
	\begin{tikzpicture}
	\draw[lime, fill=lime] (0,0)
	circle [radius=0.16]
	node[white] {{\fontfamily{qag}\selectfont \tiny ID}};
	\draw[white, fill=white] (-0.0625,0.095)
	circle [radius=0.007];
	\end{tikzpicture}
	\hspace{-2mm}
}

\newcommand{\orcidauthorA}{0000-0002-3826-7427}
\newcommand{\orcidauthorB}{0000-0003-1031-9261}
\newcommand{\orcidauthorC}{0000-0003-3810-5943}
\newcommand{\orcidauthorD}{0000-0001-6205-9479}

\newcommand{\orcidA}{\href{https://orcid.org/\orcidauthorA}{\orcidicon}}
\newcommand{\orcidB}{\href{https://orcid.org/\orcidauthorB}{\orcidicon}}
\newcommand{\orcidC}{\href{https://orcid.org/\orcidauthorC}{\orcidicon}}
\newcommand{\orcidD}{\href{https://orcid.org/\orcidauthorD}{\orcidicon}}

\usepackage{authblk}

\author[1]{Paulo S. S. dos Santos\orcidA{}}
\author[2]{João P. Mendes\orcidB{}}
\author[1,2]{José M. M. M. de Almeida\orcidC{}}
\author[1,3]{Luís C. C. Coelho\orcidD{}}

\affil[1]{INESC TEC - Institute of Systems and Computer Engineering, Technology and Science. Rua Dr. Roberto Frias, 4200-465 Porto, Portugal}
\affil[2]{Department of Physics, School of Science and Technology, University of Trás-os-Montes e Alto Douro, 5001-801 Vila Real, Portugal}
\affil[3]{Department of Physics, Faculty of Sciences, University of Porto, Rua do Campo Alegre, 4169-007 Porto, Portugal }

\begin{document}

\twocolumn[
  \begin{@twocolumnfalse}

\maketitle

\begin{abstract}
Accurate and fast calculations of localized surface plasmon resonances (LSPR) in metallic nanoparticles is essential for applications in sensing, nano-optics, and energy harvesting. Although full-wave numerical techniques such as the boundary element method (BEM) or the discrete dipole approximation (DDA) provide high accuracy, their computational cost often hinders rapid parametric studies. Here it is presented an ultrafast method that avoids solving large eigenproblems. Instead, only the dipolar component of the induced surface charge density \((\sigma_{dipolar})\) is retained through a expansion into Cartesion dipole basis, yielding a compact $3\times3$ geometric formulation that avoids full boundary-integral solves. The spectral response is obtained in a similar way, by projecting the Neumann--Poincar\'e surface operator onto the dipole subspace and evaluating a Rayleigh quotient, giving geometry-only eigenvalues again without an $N\times N$ eigenproblem. A major advantage of this method is that all geometry-dependent quantities are computed once per nanoparticle, while material dispersion and environmental changes enter only through simple algebraic expressions for the polarizability, enabling rapid evaluation across wavelengths. Retardation effects are incorporated through the modified long-wavelength approximation (MLWA), extending accuracy into the weakly retarded regime. The resulting framework provides a valuable tool for fast modelling and optimization of plasmonic nanoparticles at a significant lesser computational cost than BEM, DDA, and other standard tools.
\end{abstract}

\vspace{0.35cm}

  \end{@twocolumnfalse}
]



\section{Introduction}

Localized surface plasmon resonances (LSPR) are non-propagating collective oscillations of conduction electrons supported by metallic nanoparticles (NPs) whose characteristic dimensions are smaller than the wavelength of the incident electromagnetic radiation~\cite{Sepulveda2009}. These resonances arise from the balance between the external driving field and the Coulomb restoring forces acting on the displaced electron cloud, leading to pronounced spectral features and strong electromagnetic field enhancement both inside the nanoparticle and in its immediate surroundings~\cite{Thollar2018}. The enhanced near fields are typically confined to a few nanometers from the nanoparticle surface, making LSPR particularly sensitive to changes in the local dielectric environment~\cite{DOSSANTOS2025136936}. This extreme field confinement underpins a wide range of applications, including refractive-index-based biosensing, surface-enhanced spectroscopy, and nano-optical signal transduction~\cite{XU2021112850, dosSantos:24, pauloHybrid}.

From a theoretical perspective, the electromagnetic response of a plasmonic nanoparticle is fundamentally governed by the induced surface charge density distribution \((\sigma)\) that develops at the interface between the metal and the surrounding dielectric medium. It also determines the near-field enhancement, the far-field scattering and absorption, and the coupling of the nanoparticle to external probes such as photons or fast electrons~\cite{Sepulveda2009}. In the subwavelength regime, where retardation effects are weak, the problem reduces to an electrostatic boundary-value problem, and $\sigma$ satisfies an integral equation involving the Neumann--Poincar\'e (NP) boundary operator~\cite{6095828, FARIA2025135}. The eigenvalues of this operator encode the intrinsic plasmonic resonance conditions of the nanoparticle and depend solely on its geometry.

Several numerical approaches have been developed to solve this electrostatic boundary-value problem with high accuracy. Full-wave electrodynamic techniques, such as the boundary element method (BEM) and the boundary integral equation (BIE) method, provide detailed solutions for both near- and far-field quantities across a broad spectral range. However, these methods require solving large linear systems or eigenvalue problems whose size scales with the number of discretized surface elements~\cite{6095828, Hohenester2012}. As a consequence, they can become computationally demanding when applied to complex geometries, high aspect-ratio nanoparticles, or large parameter sweeps over wavelength and surrounding refractive index. This computational burden poses a significant limitation for applications such as plasmonic biosensing, where rapid geometry optimization and repeated evaluation of spectral shifts are often required.

At the same time, it is well established that, for typical plasmonic nanoparticles, the optical response in the visible and near-infrared is overwhelmingly dominated by the lowest-order dipolar plasmon modes~\cite{C4CS00131A}. Higher-order multipoles contribute weakly to far-field observables and become significant only for larger nanoparticles or at shorter wavelengths~\cite{10.1063/1.4914117}. This observation motivates the development of reduced-order models that focus explicitly on the dominant dipolar physics while retaining sensitivity to nanoparticle geometry and orientation. Analytical models based on ellipsoidal geometries, such as Mie-Gans theory, provide valuable insight but are restricted to highly idealized shapes and cannot be readily extended to arbitrary nanostructures encountered in realistic fabrication processes~\cite{acs.jpclett.1c03144}.

In this work,it is introduced an ultrafast electrostatic formalism that exploits the dominance of dipolar plasmon modes to enable rapid and accurate modeling of plasmonic nanoparticles with arbitrary geometry. Rather than solving the full boundary integral equation for $\sigma$, the electrostatic problem is projectd onto a Cartesian dipole subspace, yielding a compact $3\times3$ formulation that depends only on geometric moments of the nanoparticle surface. This approach allows direct reconstruction of the dipolar surface charge distribution and extraction of geometry-dependent plasmonic eigenvalues through a Rayleigh-quotient evaluation of the projected K operator, entirely avoiding large-scale matrix inversions or eigenvalue solves. Crucially, all geometry-dependent quantities are computed once per nanoparticle, while material dispersion and environmental changes enter only through simple algebraic expressions for the polarizability. Retardation effects are incorporated via the modified long-wavelength approximation (MLWA), extending the validity of the model into the weakly retarded regime relevant for elongated or anisotropic nanoparticles. In the following sections, it is presented the theoretical formulation, and demonstrate appropriate application scenarios.

\section{Theory} \label{sec:method}

The first step of the present framework is to determine the induced surface charge density \(\sigma(\mathbf{s})\) on the nanoparticle surface \(\mathbf{s}\) under external electromagnetic excitation, building a geometry-only second-moment tensor M. This step provides a fast and robust approximation to the dominant dipolar mode, suitable for near-field mapping and mode orientation. Next, the spectral response is obtained by projecting the K surface operator onto a Cartesian dipolar subspace, yielding geometry-dependent eigenvalues that correct the resonance condition without requiring a large eigenvalue solve. On the following subsections, it is detailed each of these steps.

\subsection{Surface charge density \((\sigma\))} \label{subsec:sigma_calc}

In the quasistatic regime \(ka < 0.3\), where $k = 2\pi/\lambda$ is the wavenumber in the embedding medium, and \(a\) a characteristic dimension for the nanoparticle, the electromagnetic response can be greatly simplified by retaining only a few dominant multipolar contributions, rather than solving the full spatially complex charge distribution. In this regime Maxwell's equations for a body of permittivity \(\epsilon_m\) in a homogeneous medium of permittivity \(\epsilon_d\), is reduced to a Poisson's equation. Therefore, the fundamental boundary integral equation to solve is~\cite{PhysRevB.65.115418}:

\begin{equation}
  \varGamma  \sigma(\mathbf{s}) = f(\mathbf{s}) + \int_{s} ds' F(\mathbf{s}, \mathbf{s'}) \sigma(\mathbf{s})
  \label{eq:fundamental}
\end{equation}

where

\begin{equation}
  \varGamma = 2 \pi \frac{\epsilon_d + \epsilon_m}{\epsilon_d - \epsilon_m}
\end{equation}

is the parameter that depends on the contrast between the two media. f(s) is the projection of the external field onto the interface normals, and F(s, s') is the normal derivative of the Coulomb potential kernel evaluated on the interface:

\begin{equation}
  F(s,s') = \mathbf{n}_s \cdot \nabla_s \frac{1}{|\mathbf{s}-\mathbf{s}'|}
         = \mathbf{n}_s \cdot \frac{\mathbf{s}-\mathbf{s}'}{|\mathbf{s}-\mathbf{s}'|^3}
\end{equation}

This represents the normal component of the electric field at \(\mathbf{s}\) due to a unit charge at \(\mathbf{s}'\), i.e., the Coulomb kernel acted on by
\(\mathbf{n}_s \cdot \nabla\). Here, the BEM approach converts the integral equation into a linear system which is then solved by direct matrix inversion, resulting in a \(N \times N\) linear system, which must be repeated for each \(\lambda\), and for each time the permittivity conditions change. 

Here, it is proposed a different approach. Considering that \(\sigma(\mathbf{s})\) can in general be expanded as a series of multipole moments,
\begin{equation}
  \sigma(\mathbf{s})
  = \sigma_{\text{monopole}}(\mathbf{s})
  + \sigma_{\text{dipole}}(\mathbf{s})
  + \sigma_{\text{quadrupole}}(\mathbf{s})
  + \cdots 
\end{equation}

For plasmonic nanoparticles, the monopole term can be discarded as the overall net charge is usually zero. The dipolar term is the first non-trivial and dominant term. Then, imposing a Cartesian expansion on this term:

\begin{equation}
  \sigma(\mathbf{s}) \approx B_x \, x(\mathbf{s}) + B_y \, y(\mathbf{s}) + B_z \, z(\mathbf{s}),
\end{equation}

where $x(\mathbf{s}), y(\mathbf{s}), z(\mathbf{s})$ are the Cartesian coordinates of the surface point $\mathbf{s}$, and $B_x, B_y, B_z$ are coefficients associated with the amplitudes of the dipole moment components along the respective Cartesian axes. Together, they summarize the nanoparticle's dipole moment vector and are sufficient to capture the dominant electromagnetic response. Inserting this expansion into equation (\ref{eq:fundamental}) results on its multiplication by each of the basis functions $r_i(\mathbf{s})$, where $r_i$ is the $i$-th coordinate function, i.e., $x$, $y$, or $z$, giving

\begin{align}
  \int_S r_i(\mathbf{s})\,\sigma(\mathbf{s})\,\mathrm{d}S  &= \int_S r_i(\mathbf{s})\,f(\mathbf{s})\,\mathrm{d}S \notag \\
  &\quad + \int_S \int_S r_i(\mathbf{s})\,F(\mathbf{s},\mathbf{s}')\,\sigma(\mathbf{s}')\,\mathrm{d}S \mathrm{d}S'
  \label{eq:integralForm}
\end{align}

The left-hand side represents the projection of the unknown surface charge multiplied by $r_i$, essentially the dipole moment component along the $i$-th direction due to $\sigma$. The first term on the right-hand side is the projection of the external driving function $f(\mathbf{s})$ onto $r_i$. The second term on the right-hand side represents the coupling between charge at $\mathbf{s}'$ and the point $\mathbf{s}$, over the surface, then projected onto $r_i(\mathbf{s})$. Writing equation (\ref{eq:integralForm}) into matrix form yields the operator

\begin{equation}
M_{ij} = \int_{S}\!\!\int_{S} r_i(\mathbf{s})\,F(\mathbf{s},\mathbf{s}')\,r_j(\mathbf{s}')\,dS'\,dS - \delta_{ij}\int_{S} r_i(\mathbf{s})^2\,dS
\end{equation}

Similarly, the vector $\mathbf{f}=(f_x,f_y,f_z)$ contains the projections of the external excitation $f(\mathbf{s})$ onto the basis functions:

\begin{equation}
f_i = \int_{S} r_i(\mathbf{s})\,f(\mathbf{s})\,dS
\end{equation}

Combining all, one obtains a linear algebraic system:
\begin{equation}
\begin{bmatrix}
M_{xx} & M_{xy} & M_{xz}\\
M_{yx} & M_{yy} & M_{yz}\\
M_{zx} & M_{zy} & M_{zz}
\end{bmatrix}
\begin{bmatrix}
B_x\\
B_y\\
B_z
\end{bmatrix}
=
\begin{bmatrix}
f_x\\
f_y\\
f_z
\end{bmatrix}
\end{equation}

This \(3\times 3\) matrix encodes the geometry and electromagnetic interaction kernel integrated over the surface. The off-diagonal terms $M_{ij}$ with $i\neq j$ represent coupling between different dipole directions due to anisotropic or asymmetric shapes, while the diagonal terms $M_{ii}$ represent the response along each Cartesian axis. Here, the computational bottleneck is the direct evaluation of the matrix entries \(M_{ij}\) by double summation over all face pairs, scaling as \(\mathcal{O}(N^{2})\) for \(N\) surface elements. Fortunately, the integral can be approximated using geometric moments, a valid approach when the charge or source varies slowly on each face~\cite{Gibson2021}. The key idea is to approximate the surface charge distribution inside a given element by its low-order moments around the centroid, simplifying the double integral. The approximation replaces the integral by contributions weighted around the centroid of the discretized faces:

\begin{equation}
\mathbf{M}_{ij} \approx \frac{1}{V}\sum_{\text{faces}} \left(r_i - \bar{r}_i\right)\left(r_j - \bar{r}_j\right)A
\end{equation}

where $r_i$ is the $i$-th coordinate at a given face, $A$ is the area of each face, $V$ is the total volume enclosed (or a normalization factor), and $\bar{r}_i$ is the centroid (average coordinate) of all surface elements, defined as

\begin{equation}
\bar{r_i} = \frac{1}{A}\int_{S} r_i \,\mathrm{d}S
\label{eq:surface_centroid}
\end{equation}
All position vectors are henceforth expressed in terms of centered coordinates $\mathbf{r}-\bar{\mathbf{r}}$. This approach effectively replaces computing double integrals over all pairs of faces by a single sum over faces. The final system is a manageable \(3\times 3\) linear algebraic problem. Solving this system provides the coefficients \(B_x\), \(B_y\), \(B_z\) of the induced dipolar charge distribution, encapsulating the primary electromagnetic response of the nanoparticle.

in the quasistatic limit considered here, $\mathbf{M}_{ij}$ is strictly real, being that any imaginary contributions arise from numerical quadrature errors. Next, $\mathbf{M}$ is diagonalized via

\begin{equation}
\mathbf{M}_\text{real} = \mathbf{R} \boldsymbol{\Lambda} \mathbf{R}^\top
\end{equation}

where $\boldsymbol{\Lambda} = \text{diag}(\lambda_1, \lambda_2, \lambda_3)$ contains the eigenvalues of $\mathbf{M}$ and $\mathbf{R}$ is the rotation matrix to the principal axes. The resulting quantities define effective geometric depolarization factors, which provide a convenient parametrization of the nanoparticle anisotropy and principal axes, but should not be interpreted as exact electrostatic depolarization factors as in the case for ellipsoids in the Mie-Gans theory. In this case the depolarization factors for arbitrary geometries are described as

\begin{equation}
L_\text{eff, k} = \frac{1 / \lambda_k}{\sum_{j=1}^3 1/\lambda_j}, \quad k=1,2,3
\label{eq:leff}
\end{equation}

Here, the depolarization factors are the inverse of the $\mathbf{M}$ eigenvalues because $\mathbf{M}$ is not the depolarization tensor, but a second-moment / geometric tensor with units of $length$. So, the quantity that tracks $L_\text{eff, k}$ should be anti-correlated with axis length. This normalization is performed to guarantee that \(\sum_{i} L_\text{eff, i} = 1\). 

Finally, along each principal axis $k$, the polarizability is given by:

\begin{equation}
\alpha_k = 4\pi V \frac{\epsilon_m - \epsilon_d}{\epsilon_d + L_{\text{eff},k}(\epsilon_m - \epsilon_d)}
\label{eq:polarizability}
\end{equation}

The dipole moments can finally be calculated as:

\begin{equation}
\mathbf{p} = \mathbf{R}^{\mathsf{T}} \boldsymbol{\alpha} \mathbf{R} \mathbf{E}_0
\label{eq:dipole_rotation}
\end{equation}

where \(\boldsymbol{\alpha} = \mathrm{diag}(\alpha_x,\alpha_y,\alpha_z)\), and \(\mathbf{E}_0\) is the external electric field. The induced surface charge density is reconstructed via the uniform-dipole approximation:

\begin{equation}
\sigma(\mathbf{r}) = \frac{3}{4\pi V} \mathbf{p} \cdot (\mathbf{r} - \bar{\mathbf{r}})
\end{equation}

where $\mathbf{r} - \bar{\mathbf{r}}$ are the relative position vectors from the nanoparticle centroid, evaluated at the surface element centers. This yields a dipole-like approximation to the induced surface charge distribution, constructed directly from the solved dipole moments. A key advantage of this geometric moment approach is its wavelength independence for the structural computation. Although the polarizability $\alpha_k$ formally depends on the dielectric functions $\epsilon_m(\lambda)$ and $\epsilon_d(\lambda)$, the geometric tensor $\mathbf{M}$, rotation matrix $\mathbf{R}$, and effective depolarization factors $\mathbf{L}_\text{eff}$ are computed once from the nanoparticle geometry alone and remain fixed across all wavelengths.

While this geometric reconstruction captures the dominant dipolar charge pattern, the accurate determination of the resonance condition requires accounting for the electrostatic surface operator, which is addressed in the following section via a reduced operator projection.

\subsection{Calculation of far-field quantities} \label{subsec:farfield}

Far-field optical observables such as extinction and scattering cross sections are obtained from the dipolar polarizability tensor of the nanoparticle.
Within the present framework, the polarizability is constructed from geometry-dependent electrostatic eigenvalues and effective mode volumes, and subsequently corrected for retardation effects. Once the polarizability tensor is known, all far-field quantities follow directly from standard electrodynamic relations.

\subsubsection{Dipolar polarizability from electrostatic eigenvalues}
\label{subsubsec:polarizability}

\begin{figure}[!ht]
\centering
\includegraphics[width=0.9\columnwidth]{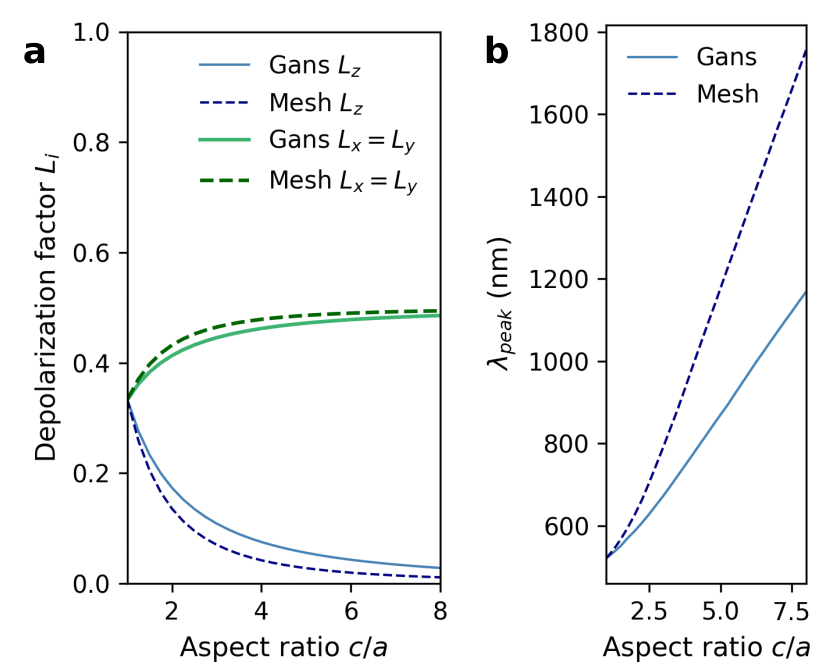}
\caption{ \(\mathbf{a}\) Comparison between the depolarization factors, calculated from \(\mathbf{M}_{geo}\) and from Mie-Gans theory for ellipsoids. \(\mathbf{b}\)
Comparison of the peak position of the extinction cross-section calculated from the base polarizability and Mie-Gans theory for ellipsoids.
}
\label{fig:basepolarizability}
\end{figure}

The usage of the quasistatic polarizability obtained solely from geometric depolarization factors can result in large errors for highly anisotropic nanoparticles. For the ellipsoid case, where an analytical description of the depolarization factors are available through the Mie-Gans theory, it is obvious the large discrepancy increasing with the elongation of the ellipsoids (See S.I. Fig.~\ref{fig:basepolarizability} $\mathbf{a}$). The major consequence of this is revealed when inserting the $L_\text{eff, k}$ factors into equation~(\ref{eq:polarizability}), where a too low, $L_\text{eff, k}$ forces the resonance condition to require a more negative $\Re[\varepsilon]$, which occurs at longer wavelengths, producing a systematic over redshift in the predicted peak, (See S.I. Fig.~\ref{fig:basepolarizability} $\mathbf{b}$). So, to accurately predict the LSPR peak position \(\lambda_{\text{peak}}\), it is essential to incorporate the correct electrostatic resonance conditions. Here it is adopted the general formulation expressed in terms of effective permittivity eigenvalues \(\varepsilon_n\) associated with each dipolar mode \(n\) as described by Shahbazyan et al.,~\cite{PhysRevA.107.L061503}:

\begin{equation}
  \alpha_k (\lambda) = V_n \frac{\varepsilon_m(\lambda) - \varepsilon_d(\lambda)}{\varepsilon_m(\lambda) - \varepsilon_n}
\label{eq:alpha_universal}
\end{equation}

In this formulation, with only the effective permittivity eigenvalue \(\varepsilon_n\) and the corresponding effective mode volume \(V_n\) it is possible to  calculate the polarizability along each principal axis and obtain a complete description of the LSPR response. The determination of \(V_n\) is explained in section~\ref{subsec:mode volume}. In the next subsection is it explained the method to calculate $\varepsilon_n$.

\subsubsection{Calculation of the permittivity eigenvalue}
\label{subsubsec:e_eff}

While the geometric-moment tensor introduced in Section~\ref{subsec:sigma_calc} provides a fast route to the \emph{driven} dipolar response of the nanoparticle under an external field, the LSPR spectral position is determined by geometry-dependent eigenvalues of the electrostatic surface operator appearing in Eq.~(\ref{eq:fundamental}). Within the boundary-element formulation, this operator corresponds to the Neumann--Poincar\'e operator (K), which acts on $\sigma(\mathbf{s})$  as ~\cite{JUNG2023103951}

\begin{equation}
(K\sigma)(\mathbf{s})
=
\frac{1}{4\pi}
\int_{S}
\frac{(\mathbf{s}-\mathbf{s}')\cdot \hat{\mathbf{n}}(\mathbf{s})}
{|\mathbf{s}-\mathbf{s}'|^{3}}
\,
\sigma(\mathbf{s}')
\,\mathrm{d}S'
\label{eq:NP_operator}
\end{equation}

where $\hat{\mathbf{n}}(\mathbf{s})$ denotes the outward unit normal vector at the surface point $\mathbf{s}$. Its eigenvalues encode the intrinsic resonance conditions of the nanoparticle independently of excitation and material dispersion. However, a direct computation of the K operator eigenvalues requires solving an $N\times N$ eigenvalue problem. Therefore, in the same spirit of the above expansion into Cartesian coordinates, $\sigma(\mathbf{s})$ is again written as a linear combination of the Cartesian coordinate functions evaluated on the nanoparticle surface

\begin{equation}
\sigma(\mathbf{s}) = \sum_{i=1}^{3} c_i\, r_i(\mathbf{s})
\end{equation}

with $r_i \in \{x,y,z\}$. Within this reduced subspace, the K eigenvalue problem $K\sigma = \kappa \sigma$ is enforced in a weak (Galerkin) sense by requiring that the residual be orthogonal to the same dipole basis. This projection leads to a generalized eigenvalue problem of the form

\begin{equation}
T\,\mathbf{c} = \kappa_n\, G\, \mathbf{c}
\label{eq:projected_eigenproblem}
\end{equation}

where the matrices $G$ and $T$ encode, respectively, the inner products of the dipole basis functions and the action of the K operator within this subspace. Explicitly, they are defined as

\begin{align}
G_{ij} &= \int_S r_i(\mathbf{s})\, r_j(\mathbf{s})\, \mathrm{d}S \\
T_{ij} &= \int_S r_i(\mathbf{s})\, (K r_j)(\mathbf{s})\, \mathrm{d}S
\end{align}

In discretized form, both matrices are evaluated by single summations over surface elements and therefore scale linearly with the number of faces. Solving Eq.~(\ref{eq:projected_eigenproblem}) yields up to three eigenvalues $\kappa_n$, corresponding to the optimal dipolar approximations of the true K eigenvalues supported by the nanoparticle geometry.

Solving the generalized eigenvalue problem of Eq.~(\ref{eq:projected_eigenproblem}) yields the stationary values of the Rayleigh quotient associated with the Neumann--Poincar\'e operator

\begin{equation}
\kappa = \frac{\langle \sigma, K \sigma \rangle}{\langle \sigma, \sigma \rangle}
\end{equation}

where the inner product is defined over the nanoparticle surface. The restriction of this variational problem to the dipole subspace spanned by the Cartesian coordinate functions ensures that the resulting eigenvalues $\kappa_n$ correspond to the optimal dipole-like approximations of the true NP eigenvalues supported by the nanoparticle geometry. Each projected eigenvalue $\kappa_n$ is associated with an effective permittivity eigenvalue $\varepsilon_n$ through the standard K resonance condition

\begin{equation}
\varepsilon_n = -\varepsilon_d \frac{1 + 2\kappa_n}{1 - 2\kappa_n}
\label{eq:eps form k}
\end{equation}

As $\kappa_n$ depend solely on the nanoparticle geometry, only needs to be computed once, allowing $\varepsilon_n$ to be calculated rapidly for any changing in the permittivity of the nanoparticle or surrounding medium. In contrast to the geometric-moment tensor, which enables ultrafast evaluation of the \emph{driven} optical response for arbitrary dispersive material models, the K eigenvalues extracted here provide the intrinsic geometric parameters that determine the spectral location of the LSPR. Together, these two ingredients allow a clear separation between geometry-dependent characterization and excitation-dependent response, forming the basis of the ultrafast framework developed in this work.

\subsection{Calculation of effective mode volume} \label{subsec:mode volume}

Once the effective permittivity eigenvalue $\varepsilon_n$ of the dominant dipolar mode has been determined via the projected K operator (Section~\ref{subsubsec:e_eff}), an effective mode volume can be constructed following the universal polarizability formalism of Ref.~\cite{PhysRevA.107.L061503}. In the strict theory, the mode volume $V_n$ is defined from the corresponding electromagnetic eigenfields. Since the present method does not resolve the full eigenfield distribution, it is instead introduced a \emph{dipole-subspace effective mode volume} that captures the dominant dipolar oscillator strength using only geometry and the dipole-like K modes. The effective mode-volume tensor is defined as

\begin{equation}
\mathbf{V}_n^{\mathrm{eff}} = \frac{V}{4\pi} \left| \frac{\varepsilon_n}{\varepsilon_d} - 1 \right| \mathbf{s}_n,
\label{eq:Veff_tensor}
\end{equation}

where $V$ is the physical nanoparticle volume and $\varepsilon_d$ is the dielectric permittivity of the surrounding medium. Equation~(\ref{eq:Veff_tensor}) is the tensorial counterpart of the scalar expression $V_n = V_m |\chi'| s_n$ appearing in Eq.~(2) of Ref.~\cite{PhysRevA.107.L061503}, and separates the material-dependent factor (contained in $\varepsilon_n$) from the purely geometric, dimensionless factor encoded in $\mathbf{s}_n$, which encodes how the total dipolar oscillator strength is distributed among orthogonal spatial directions. This parameter can be directly determined from the dipole-like Neumann--Poincar\'e modes obtained by projecting the K operator onto the Cartesian dipole subspace $\{x(\mathbf{s}),y(\mathbf{s}),z(\mathbf{s})\}$. Denoting these three dipole-subspace eigenmodes by their surface charge distributions $\sigma_n(\mathbf{s})$ (normalized within the same discretized inner product), the associated dipole moments are computed

\begin{equation}
\mathbf{p}_n \propto \int_S (\mathbf{r}-\bar{\mathbf{r}})\,\sigma_n(\mathbf{s})\,\mathrm{d}S,
\label{eq:pn_def}
\end{equation}

which define both the polarization direction of each dipole-like mode and its coupling strength to an incident field. For a given excitation polarization $\hat{\mathbf{e}}$ (unit vector), each dipole-subspace mode is assigned a non-negative coupling weight

\begin{equation}
w_n \propto \|\mathbf{p}_n\|^2,
\qquad \sum_n w_n = 1,
\label{eq:weights}
\end{equation}

The weights $w_n$ provide an excitation-independent measure of dipolar oscillator strength within the dipole subspace. Next, constructing the normalized shape tensor as a coupling-weighted sum of rank-one projectors onto the modal dipole directions,

\begin{equation}
\mathbf{s}_n = \frac{1}{\mathrm{Tr}(\mathbf{S})}\, \mathbf{S},
\qquad
\mathbf{S}=\sum_{n=1}^{3} w_n\,\hat{\mathbf{u}}_n\hat{\mathbf{u}}_n^\top,
\label{eq:sn_projectors}
\end{equation}

where $\hat{\mathbf{u}}_n=\mathbf{p}_n/\|\mathbf{p}_n\|$ is the unit dipole direction of mode $n$. With this definition, $\mathbf{s}_n$ is symmetric, positive semidefinite, dimensionless, and depends only on geometry through the dipole-subspace K modes. When the excitation is aligned with a single dominant dipole-like mode, $\mathbf{s}_n$ approaches the corresponding projector $\hat{\mathbf{u}}\hat{\mathbf{u}}^\top$. For isotropic nanoparticles, the three dipole modes are degenerate and $\mathbf{s}_n$ reduces to an isotropic tensor.

\begin{figure}[!ht]
\centering
\includegraphics[width=0.9\columnwidth]{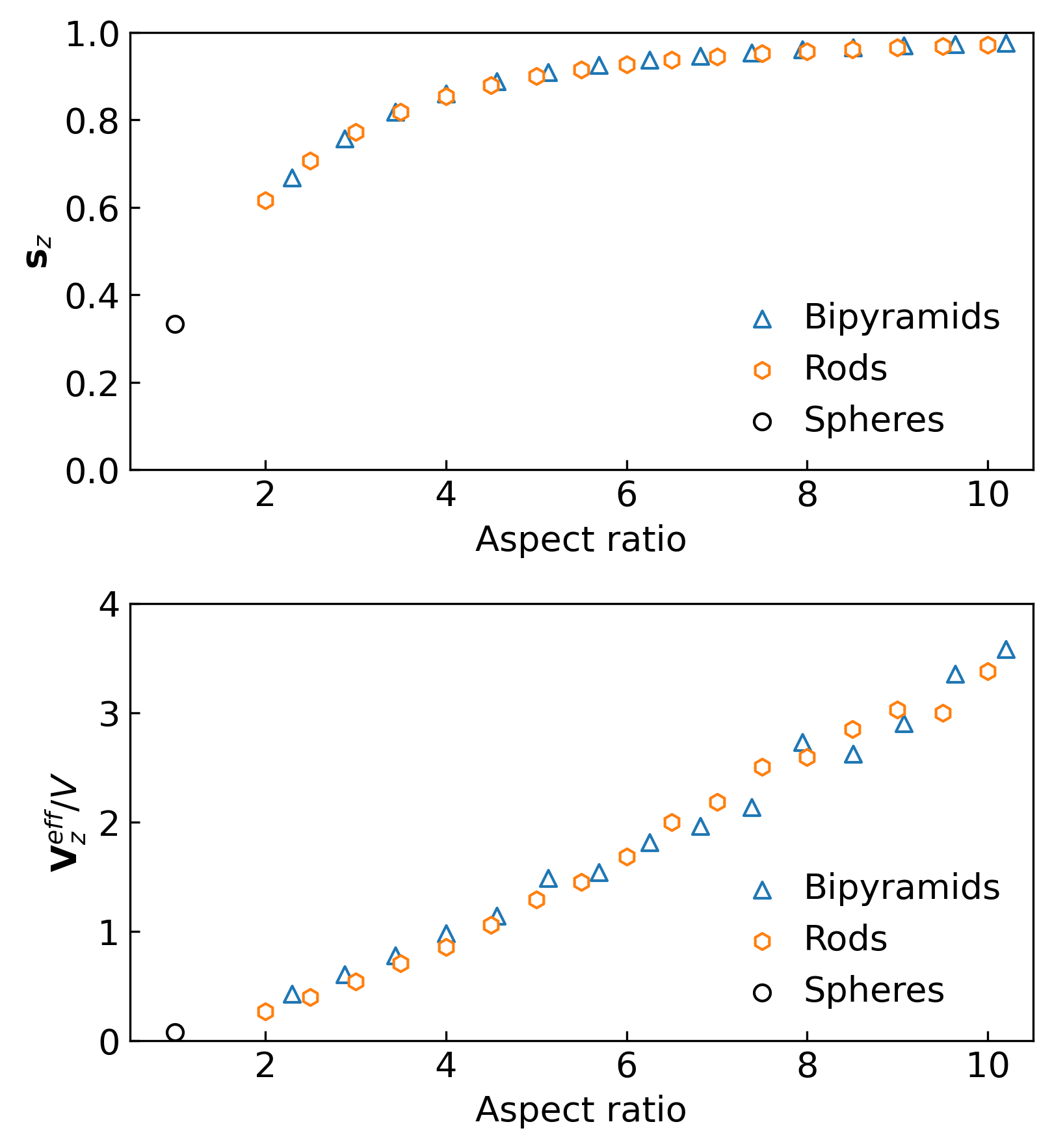}
\caption{
Normalized shape tensor $\mathbf{s}_n$ and normalized mode volumes \( \mathbf{V}_n^{eff} / V \) for a gold nanorods, bipyramids and spheres with the longer axis aligned with the external field, along the z-axis.
}
\label{fig:s_n}
\end{figure}

Diagonalizing $\mathbf{V}_n^{\mathrm{eff}}$ yields three principal effective mode volumes $(V_x,V_y,V_z)$ and a corresponding rotation matrix to the principal basis. These volumes are subsequently used to build the quasistatic polarizability in the principal frame. Importantly, $\mathbf{s}_n$ defined by Eq.~(\ref{eq:sn_projectors}) avoids conflating oscillator strength with depolarization factors. Furthermore, as the effective mode volume defined in Eq.~\eqref{eq:Veff_tensor} is a response-weighted quantity and is not bounded by the physical nanoparticle volume; values exceeding unity reflect the strong extension of plasmonic near fields into the surrounding dielectric (See S.I. Fig.~\ref{fig:s_n}).

\subsubsection{Retardation corrections to the quasistatic polarizability} \label{subsubsec:mlwa}

The quasistatic polarizability tensor constructed from the effective permittivity eigenvalues $\varepsilon_n$ accurately captures the electrostatic response of subwavelength nanoparticles. However, as the nanoparticle size or aspect ratio increases, retardation effects become non-negligible, leading to dynamic depolarization and radiative damping that are not accounted for in the purely quasistatic description. To incorporate these effects while preserving the computational efficiency of the present formalism, the modified long-wavelength approximation (MLWA) is used~\cite{acs.jpcc.0c09774}. Within MLWA, the polarizability along each principal dipolar direction $\kappa$ is corrected according to

\begin{equation}
\alpha_n^{\mathrm{MLWA}}=
\frac{\alpha_n^{(0)}}
{1-\dfrac{k^2}{a_n}\alpha_n^{(0)}-\dfrac{2i}{3}k^3\alpha_n^{(0)}}
\label{eq:mlwa}
\end{equation}

where $\alpha_n^{(0)}$ is the quasistatic polarizability given by Eq.~(\ref{eq:polarizability}). The term $\kappa^2 \alpha_k^{0} / a_k $ is the dynamic depolarization correction, $a_k$ is taken as the semi-extent along principal axis $\kappa$. This term being a real quantity, it is responsible to shift the band towards longer wavelengths. The imaginary term proportional to $\kappa^3$ accounts for radiative damping, and produces a broadening of the band. Together, these corrections extend the validity of the quasistatic polarizability into the weakly retarded regime \(\kappa a \lesssim 0.7\).

\subsubsection{Far-field observables}
\label{subsubsec:farfield_observables}

Once the MLWA-corrected polarizability tensor $\boldsymbol{\alpha}$ is obtained, far-field optical observables follow directly from standard dipole scattering
theory. In the dipolar limit, the extinction cross section is given by the optical theorem~\cite{Bohren1998}

\begin{equation}
C_{\mathrm{ext}} = \frac{4\pi}{k^2}
\operatorname{Im}
\!\left[
\operatorname{Tr}(\boldsymbol{\alpha})
\right]
\label{eq:extinction}
\end{equation}

where $k = 2\pi n_d / \lambda$ is the wavenumber in the surrounding medium. The scattering cross section associated with electric dipole radiation is

\begin{equation}
C_{\mathrm{sca}} = \frac{k^4}{6\pi}
\operatorname{Tr}
\!\left(
\boldsymbol{\alpha}\,
\boldsymbol{\alpha}^\dagger
\right)
\label{eq:scattering}
\end{equation}

where the dagger denotes the Hermitian conjugate. This expression accounts for radiative losses arising from dipole emission into the far field. The absorption cross section is obtained from energy conservation as the difference between extinction and scattering,

\begin{equation}
C_{\mathrm{abs}} = C_{\mathrm{ext}} - C_{\mathrm{sca}}
\label{eq:absorption}
\end{equation}

and quantifies the power dissipated within the nanoparticle due to material losses. Equations~(\ref{eq:extinction})--(\ref{eq:absorption}) provide a complete description of the far-field optical response of the nanoparticle within the dipolar approximation, and can be evaluated efficiently for arbitrary dispersive material models once the polarizability tensor is known.

The angular distribution of the scattered radiation is obtained from the far-field intensity of an oscillating electric dipole. The differential
scattering cross section in the observation direction $\hat{\mathbf{r}}$ is given by

\begin{equation}
\frac{\mathrm{d}C_{\mathrm{sca}}}{\mathrm{d}\Omega}
=
\frac{k^4}{16\pi^2}
\left|
\hat{\mathbf{r}}
\times
\bigl(
\hat{\mathbf{r}}
\times
\boldsymbol{\alpha}\mathbf{E}_0
\bigr)
\right|^2
\label{eq:differential_scattering}
\end{equation}

where $\mathbf{E}_0$ denotes the incident electric field amplitude. This expression provides direct access to the full scattering pattern and allows angular-resolved observables to be computed without additional numerical solution of Maxwell's equations.

\section{Results} \label{sec:results}

\subsection{Validation of dipolar surface charge reconstruction} \label{sec:results sigma}

\begin{figure}[t]
  \centering
  \includegraphics[width=0.95\columnwidth]{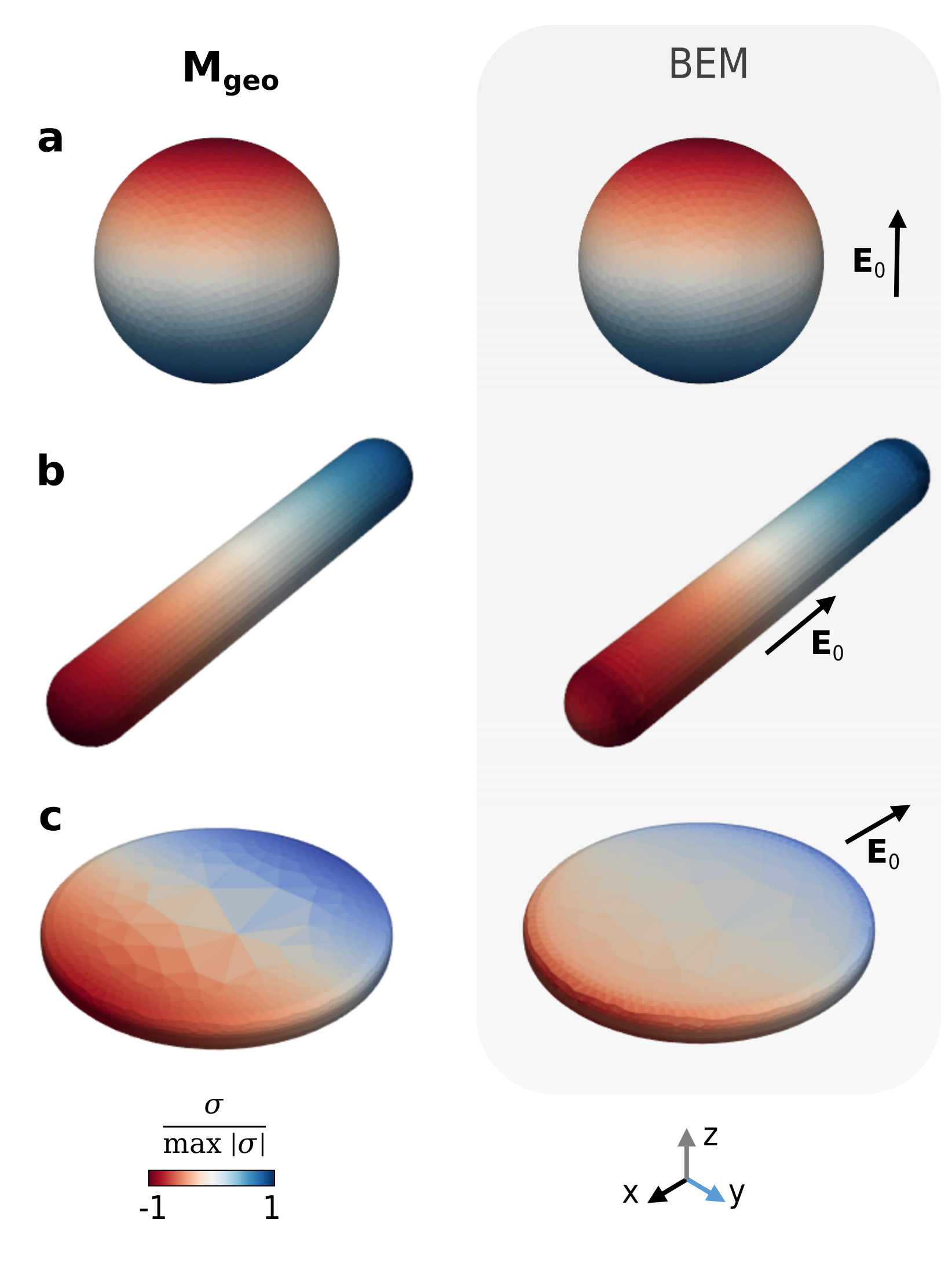}
  \caption{Normalized surface charge density comparison between the geometric-moment approach and the BEM for three nanoparticle geometries: \( \mathbf{a} \) Nanosphere (r = 10 nm) with the external field along the z-axis; \( \mathbf{b} \) Nanorod (w = 10 nm, l = 60 nm) with the external field applied along the x-axis; \( \mathbf{c} \) nanodisk (r = 10 nm, h = 2 nm) with the field along the x-axis.}
  \label{fig:results sigma comparison}
\end{figure}

The ability of the dipolar geometric-moment approach to reconstruct the dominant dipolar surface charge distribution is assessed by direct comparison with a full BEM solution for three representative nanoparticle geometries. In all cases, the particles are illuminated by a uniform external electric field, and the resulting surface charge densities are normalized to their respective maximum absolute values to facilitate direct comparison of spatial patterns. The normalized 
\(\sigma\) distributions for the geometric-moment method and the BEM are shown in Fig.~\ref{fig:results sigma comparison}.

For a nanosphere of radius 10 nm excited along z, the geometric-moment method recovers the expected dipolar with the correct orientation and antisymmetric charge distribution, in excellent agreement with the BEM solution. A similar level of agreement is obtained for a nanorod (width 10 nm, length 60 nm) with the external field applied along the rod axis, where the reconstructed dipole correctly reproduces the enhanced charge accumulation at the rod ends. For the nanodisk (radius 10 nm, height 2 nm) illuminated along the x direction, the geometric-moment method again captures the global dipolar pattern, including the correct sign distribution and large-scale spatial variation across the disk faces.

Deviations between the two approaches are primarily confined to localized regions near sharp edges, such as the nanodisk rim, where higher-order multipolar contributions become significant in the BEM solution. These higher-order modes also determine the steeper gradients of the surface charge density near the borders compared to the smoother variation predicted by the purely dipolar reconstruction. Despite these localized differences, the overall agreement demonstrates that the geometric-moment approach reliably recovers the dominant dipolar surface charge distribution across a broad set of particle shapes, while systematically neglecting higher-order terms such as quadrupolar contributions responsible for edge-localized charge accumulation.

\subsection{Near-field mapping }
\label{subsec:nearfield_mapping}

\begin{figure}[t]
  \centering
  \includegraphics[width=1\columnwidth]{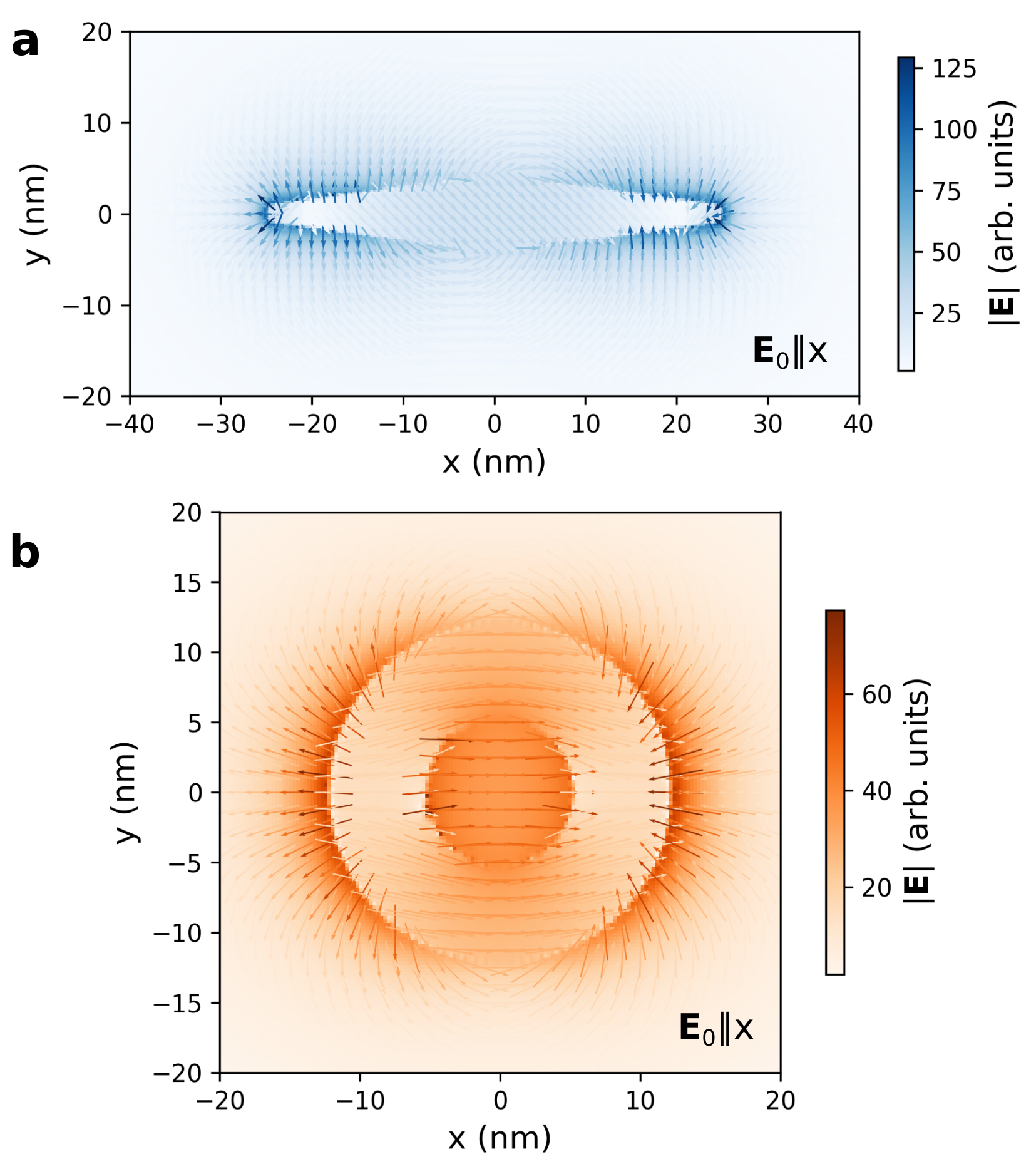}
  \caption{Near-field magnitude (color map) and vector field (arrows) computed in the xy-plane for a gold nanobipyramid (\(\mathbf{a}\)) and a gold nanoring immersed in water (\(\mathbf{b}\)), under linearly polarized excitation with the incident electric field oriented along the x-axis. The fields are reconstructed from $\sigma(s)$ and evaluated within the quasistatic approximation. The arrows indicate the local field direction, while the color scale represents the field magnitude $|\mathbf{E}|$ in arbitrary units.}
  \label{fig:results nearfield}
\end{figure}

After determining \(\sigma\), the corresponding electric field at an observation point $\mathbf{r}$ follows from $\mathbf{E}(\mathbf{r})=-\nabla\Phi(\mathbf{r})$, as:

\begin{equation}
\mathbf{E}(\mathbf{r}) = \frac{1}{4\pi\varepsilon_d} \int_{\partial\Omega} \sigma(\mathbf{s})\, \frac{\mathbf{r}-\mathbf{s}}{|\mathbf{r}-\mathbf{s}|^{3}}\,dS
\label{eq:E_near}
\end{equation}

where \(\mathbf{s}\) denotes the surface points. As an example, it is computed the optical near field in the xy-plane passing through the center of two representative geometries: a gold nanobipyramid and a gold nanoring, both embedded in water. The excitation is linearly polarized along the x-axis, ensuring dominant coupling to the longitudinal dipolar mode in both structures. In Fig.~\ref{fig:results nearfield} \(\mathbf{a}\) it is mapped the magnitude of the produced electric field, as well as its vectorial components for the nanobipyramid. The field magnitude exhibits strong localization at the two opposing tips. This behavior reflects the well-known lightning-rod effect associated with sharp features. The field vectors point predominantly along the nanoparticle axis, with rapid decay away from the tips, consistent with a dipole-like $1/r^3$ near-field dependence. In contrast, the nanoring geometry in Fig.~\ref{fig:results nearfield} \(\mathbf{b}\) displays a the field enhancement spread along the inner and outer rims of the ring, forming a continuous annular hot region. The vector field reveals circulating patterns around the ring cross-section, indicative of a delocalized dipolar charge distribution constrained by the hollow geometry. Despite these differences, the overall response remains dipolar in character, with the strongest field components aligned with the incident polarization.

Importantly, both near-field maps are obtained without introducing additional fitting parameters or solving separate field equations: they follow directly from the same dipole-subspace K surface charges that govern the extinction spectra. This demonstrates that the $M_{geo}$ alone provides a unified and self-consistent description of far-field observables and nanoscale near-field distributions. The comparison between nanobipyramid and nanoring further highlights how nanoparticle geometry controls not only resonance wavelengths but also the spatial localization and topology of plasmonic near fields under identical excitation conditions.

\subsection{Geometry-dependent dipolar eigenvalues across representative nanoparticle shapes}
\label{sec:results_eigenvalue_perm}

The geometry dependence of the dipolar eigenvalues extracted from the dipole-subspace projection of the \(\mathbf{K}\) operator is now studied. Unlike effective depolarization factors, these eigenvalues are intrinsic spectral quantities of the electrostatic surface operator and provide a unified, geometry-only description of dipolar resonances across different nanoparticle morphologies.

It is important to emphasize the distinct roles of the dipolar K eigenvalues $\kappa_n$ and the associated permittivity eigenvalues $\varepsilon_n$. The quantities $\kappa_n$ are purely geometric invariants of the nanoparticle surface and depend exclusively on the nanoparticle shape. In contrast, the permittivity eigenvalues $\varepsilon_n$, which enter the polarizability expression, are obtained by combining the geometric eigenvalues with the embedding medium permittivity through Eq.~(\ref{eq:eps form k}). As a consequence, changes in the surrounding medium shift the resonance condition by modifying $\varepsilon_n$ while leaving the underlying geometric eigenvalues $\kappa_n$ unchanged. This separation allows the geometric properties of the nanoparticle to be computed once and reused for arbitrary dielectric environments, with environmental effects entering solely through a simple rescaling of the permittivity eigenvalues. 

\begin{figure}[!ht]
\centering
  \includegraphics[width=1\columnwidth]{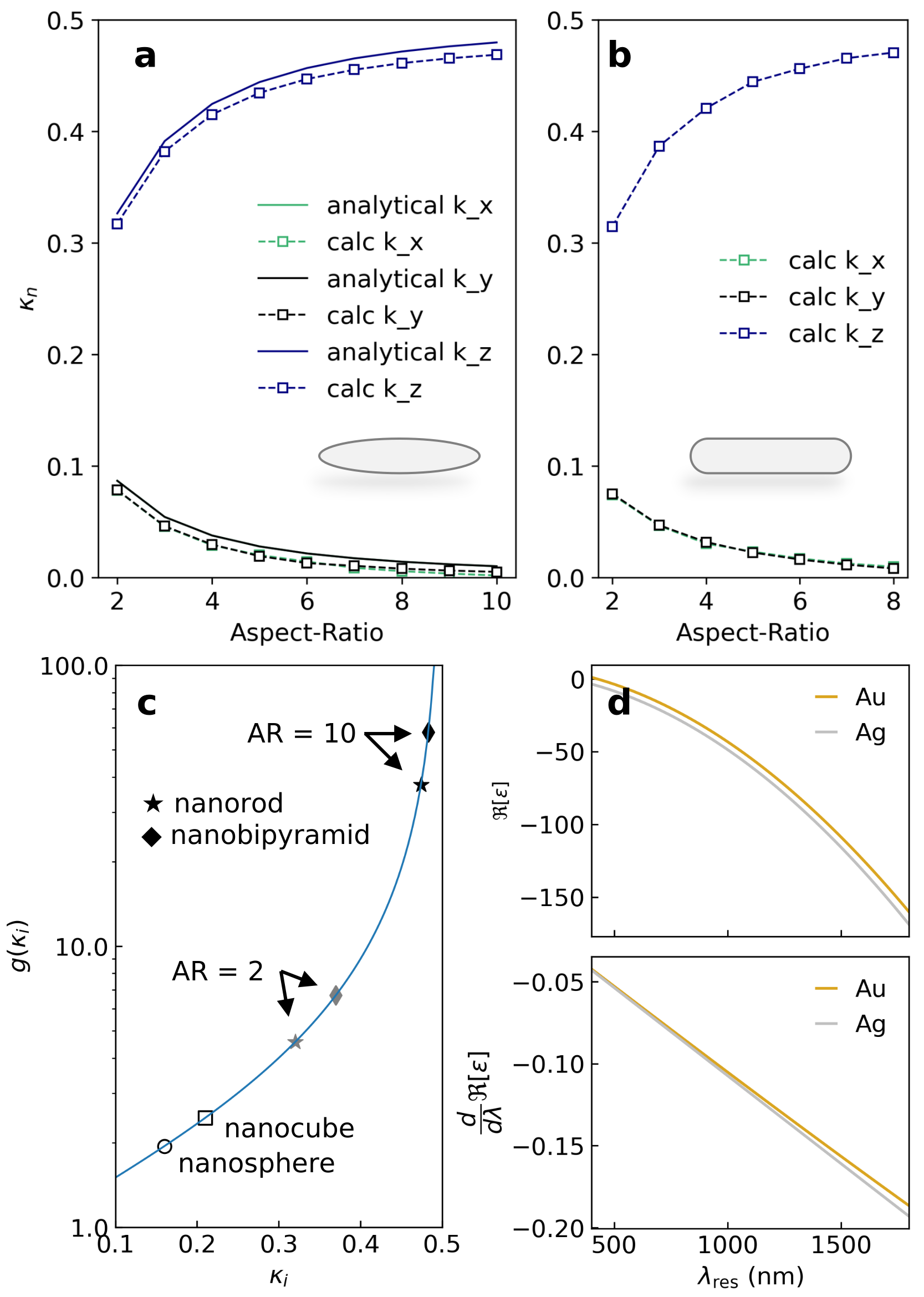}
  \caption{Geometry-dependent dipolar K eigenvalues $\kappa_n$ as a function of aspect ratio for representative nanoparticle shapes. 
  \(\mathbf{a}\) Prolate ellipsoids: comparison between analytical eigenvalues obtained from the classical depolarization factors (solid lines) and eigenvalues calculated from the dipole-subspace K projection (symbols). 
  \(\mathbf{b}\) Nanorods with spherical caps: calculated dipolar eigenvalues obtained from the K projection,
  \(\mathbf{c}\) Geometric amplification factor $g_n$ as a function of the dipolar eigenvalue $\kappa_n$.
  The nonlinear increase of $g_n$ as $\kappa_n \to 1/2$ highlights the strong enhancement of the optical response for geometries supporting large dipolar eigenvalues.
  Representative values for a sphere ($\kappa=1/6$), nanocube ($\kappa\approx0.21$), nanobipyramids, and nanorods of aspect ratios of 2 and 10 are indicated.
  \(\mathbf{d}\) Comparison between the real permittivity components (top panel), and their derivatives (bottom panel) for Au and Ag. The permittivity data is obtained from the Drude model fitted to experimental data from Johnson and Christy.}
  \label{fig:kappa_shapes}
\end{figure}

Considering prolate ellipsoids with semi-axes $(a,b,c)$, the classical depolarization factors $L_i$ $(i=x,y,z)$ satisfy $L_x+L_y+L_z=1$ and can be expressed as

\begin{equation}
L_i = \frac{abc}{2}\int_0^{\infty}\frac{\mathrm{d}s}{(s+a_i^2)\sqrt{(s+a^2)(s+b^2)(s+c^2)}},
\end{equation}

where $a_i\in\{a,b,c\}$. For prolate spheroids ($a=b<R<c$), these expressions reduce to closed-form formulas in terms of the eccentricity, and yield one longitudinal and two degenerate transverse depolarization factors. Within the K formalism, the corresponding dipolar eigenvalues are related to the depolarization factors through \(\kappa_i = 0.5 - L_i \), providing an exact analytical reference to the calculated \(\kappa_n\) from \(\mathbf{K}\). This is representative in Fig.~\ref{fig:kappa_shapes}\(\mathbf{a}\), where an excellent agreement is observed over the full range of aspect ratios, with one eigenvalue increasing monotonically as the nanoparticle elongates (longitudinal mode) and the remaining two decreasing and remaining nearly degenerate (transverse modes). This result validates the projected K approach against known analytical solutions.

Next, nanorods with spherical caps, for which no exact analytical solution exist, the calculated \(\kappa_n\) are shown in Fig.~\ref{fig:kappa_shapes}\(\mathbf{b}\) as a function of aspect ratio. As in the ellipsoidal case, one eigenvalue separates from the other two and increases with elongation, corresponding to the longitudinal dipolar mode, while the transverse modes decrease and remain nearly degenerate. The smooth evolution of the eigenvalues demonstrates that the dipole-subspace K projection yields well-defined and physically meaningful geometric eigenvalues even beyond analytically solvable shapes.

The connection between the dipolar K eigenvalues and refractive-index sensitivity can be made explicit from the resonance condition. In the quasistatic limit, the dipolar LSPR wavelength $\lambda_{\mathrm{res}}$ is determined by the matching condition

\begin{equation}
\Re\{\varepsilon_{m}(\lambda_{\mathrm{res}})\} = -\,\varepsilon_d\, \frac{1+2\kappa_z}{1-2\kappa_z}
\label{eq:res_condition}
\end{equation}

To make explicit how the geometric eigenvalues govern the optical response, it is introduced the dimensionless geometric factor \(g(\kappa_n)\):

\begin{equation}
g(\kappa_z)\equiv\frac{1+2\kappa_z}{1-2\kappa_z}
\end{equation}

which enters directly into the quasistatic resonance condition. The explicit relation between \(\kappa_n \) and \(g(\kappa_n)\) is displayed in Fig.~\ref{fig:kappa_shapes} \(\mathbf{c}\). The pronounced nonlinear increase of \(g(\kappa_n)\) as \(\kappa_n \rightarrow 0.5\) reveals the strong geometric amplification associated with elongated or sharp-featured nanoparticles. Equation~(\ref{eq:res_condition}) can be used to explicitly describe the refractive-index sensitivity of an arbitrary nanoparticle with contributions of geometry and material as (See S.I.~\ref{si_ris_derivation}) then be written compactly as

\begin{equation}
\frac{d\lambda_{\mathrm{res}}}{dn_d} = -\frac{2 n_d\, g(\kappa_z)}{\left.\dfrac{d}{d\lambda}\Re\{\varepsilon_{m}(\lambda)\}\right|_{\lambda=\lambda_{\mathrm{res}}}}
\label{eq:ris_explicit1}
\end{equation}

For noble metals such as Au and Ag, the denominator is minimized in the visible--near-IR range, where $\Re\{\varepsilon_{m}(\lambda)\}$ varies comparatively slowly (See Fig.~(\ref{fig:kappa_shapes}) \(\mathbf{d}\)), thereby increasing the material-dependent prefactor in Eq.~(\ref{eq:ris_explicit1}). Noting that as $\kappa_z\rightarrow 1/2$ the quasistatic description becomes increasingly sensitive to radiative damping and dissipative losses, so maximizing RIS alone does not necessarily maximize the sensing figure of merit.

\subsection{Extinction spectra: quasistatic versus MLWA-corrected dipole response}
\label{sec:results_extinction}

\begin{figure}[t]
\centering
\includegraphics[width=1\columnwidth]{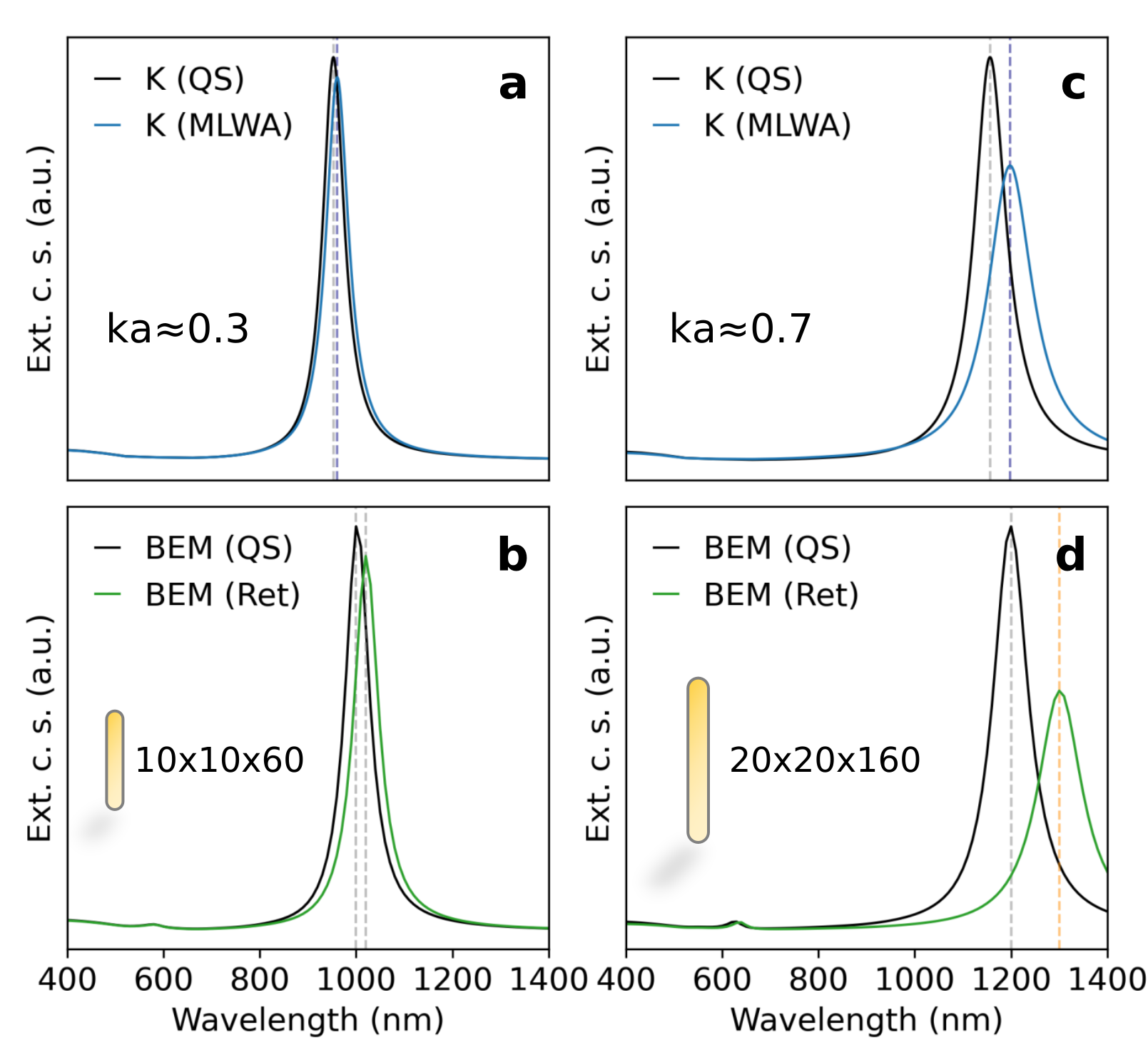}
\caption{Extinction spectra of gold nanorods in water ($\varepsilon_d = 1.77$) for excitation polarized along the long axis for a $10\times10\times60$~nm nanorod with this method (\(\mathbf{a}\)) and with BEM (\(\mathbf{b}\)). Similarly for $20\times20\times160$~nm nanorod with this method (\(\mathbf{c}\)) and with BEM
(\(\mathbf{d}\)).}
\label{fig:extinction_rods}
\end{figure}

The spectral accuracy of the framework by comparing extinction spectra computed in the quasistatic limit is compared with, and without, radiative and depolarization corrections against BEM. Considering two gold nanorods embedded in water ($\varepsilon_d = 1.77$), with dimensions $10\times10\times60$~nm and $20\times20\times160$~nm, respectively. These geometries span a regime from moderately elongated nanoparticles, for which quasistatic approximations are often adequate (\(\kappa a \approx 0.3\)), to more elongated nanoparticles, into the limit of the weakly retarded regime (\(\kappa a \approx 0.7\)). In both cases, the incident field is polarized along the nanorod long axis in order to excite the dominant longitudinal dipolar mode. In Fig.~\ref{fig:extinction_rods} \(\mathbf{a}\) is shown the extinction spectra with and without MLWA corrections for the $10\times10\times60$~nm nanorod. In this regime, the MLWA correction is minimal, producing a redshift of 7.7 nm, a full-width at half maximum (FWHM) broadening of 2.1 nm and a peak intensity decrease around 5\%. In comparison, the BEM results (See Fig.~\ref{fig:extinction_rods} \(\mathbf{b}\)) in the quasistatic regime show a band at longer wavelengths (\(\Delta \lambda = 49 nm\)) and the same FWHM. The fully retarded BEM solution further redshifts the band by an additional 20 nm and broadens the linewidth by 10 nm, indicating that retardation effects are present but remain weak. 

For the $60\times60\times240$~nm nanorod (See Fig.~\ref{fig:extinction_rods} \(\mathbf{c}\)), the MLWA demonstrate a strong correction to the quasistatic regime, producing a redshift 42 nm, a FWHM broadening of 31 nm, and a peak intensity decrease of nearly 30\%. In contrast, the retarded BEM results (See Fig.~\ref{fig:extinction_rods} \(\mathbf{d}\)) show an even stronger correction, highlighting the breakdown of the weakly retarded regime and the validity of the MLWA.«

For compact geometries such as nanospheres, the quasistatic and MLWA-corrected dipole responses are nearly indistinguishable across the spectral range considered, and both closely match full electrodynamic solutions. Although the examples presented here focus on canonical rod geometries, incorporating MLWA corrections into extinction cross-section calculations is entirely geometry-agnostic.

\subsection{Dielectric coatings}
\label{subsec:biosensor_optimization}

As a simple, experimentally relevant application of the proposed dipolar framework, consider a gold nanorod uniformly coated with a thin dielectric polymer layer, representative of common surface-functionalization strategies such as poly(vinylpyrrolidone) (PVP). Such coatings are routinely used to enhance colloidal stability and provide chemical anchoring sites, while also altering the local dielectric environment sensed by the plasmonic mode.

Within this approach, the nanoparticle geometry and its associated dipolar eigenvalues remain fixed, while the coating influences the response solely by modifying the surrounding dielectric environment. For dielectric layers with thicknesses comparable to the field decay length, the effective refractive index sampled by the plasmonic mode falls between that of water and the polymer, making an accurate description of the decay profile essential. To capture the spatial behavior of the field in a general form, an analytical model for the radial decay of the electric-field magnitude \( |\mathbf{E}(r)| \) is introduced using a power-law expression

\begin{equation}
|\mathbf{E}(r)| = \frac{A}{(r + r_0)^n}
\label{eq. fit}
\end{equation}

The choice of a power-law decay is physically motivated by the nature of electromagnetic fields generated by localized charge distributions in electrostatics. Here, the \(A\) coefficient represents the strength (amplitude) of the electric field at distances very close to the nanoparticle surface. Physically, it is expected to scale with the charge distribution and curvature of the nanoparticle surface. \(r_0\) is effectively an offset radius that accounts for the nanoparticle size. Physically, it should scale approximately with the characteristic size of the nanoparticle. The exponent n reflects how quickly the field decays spatially and depends on geometric anisotropy, e. g., when the intense regions of a dipolar resonance are located very closely. If \( n \approx 2\) the system resembles a monopole (point charge), when \( n	\approx 3 \) the system approaches a dipole-dominated decay, and for \(n < 2\) it presents a flat/distributed sheet behavior.

Using the power-law near-field decay profile in Eq.~(\ref{eq. fit}), the effective refractive index sampled by the plasmonic mode is defined as a weighted average of the surrounding dielectric environment,

\begin{equation}
n_{\mathrm{eff}} = 
\frac{
\displaystyle \int_{0}^{\infty} \eta(z)\, \frac{r_0^{\,n}}{(z+r_0)^{n}} \, dz
}{
\displaystyle \int_{0}^{\infty}
\frac{r_0^{\,n}}{(z+r_0)^{n}} \, dz
}
\label{eq:neff_powerlaw}
\end{equation}

where $\eta(z)$ is the refractive index profile normal to the nanoparticle surface, and $r_0$ and $n$ are geometry-dependent parameters extracted from the computed near-field distribution. The weighting function is normalized by construction, ensuring that $n_{\mathrm{eff}}$ represents a true effective refractive index bounded by the minimum and maximum values of $\eta(z)$.

The prefactor $r_0^{\,n}$ is chosen such that the normalized field intensity equals unity at the nanoparticle surface, i.e.\ $\left.r_0^{\,n}/(z+r_0)^n\right|_{z=0}=1$, while the denominator enforces proper normalization of the power-law decay profile over the semi-infinite dielectric environment.

\begin{figure}[!ht]
\centering
\includegraphics[width=1\columnwidth]{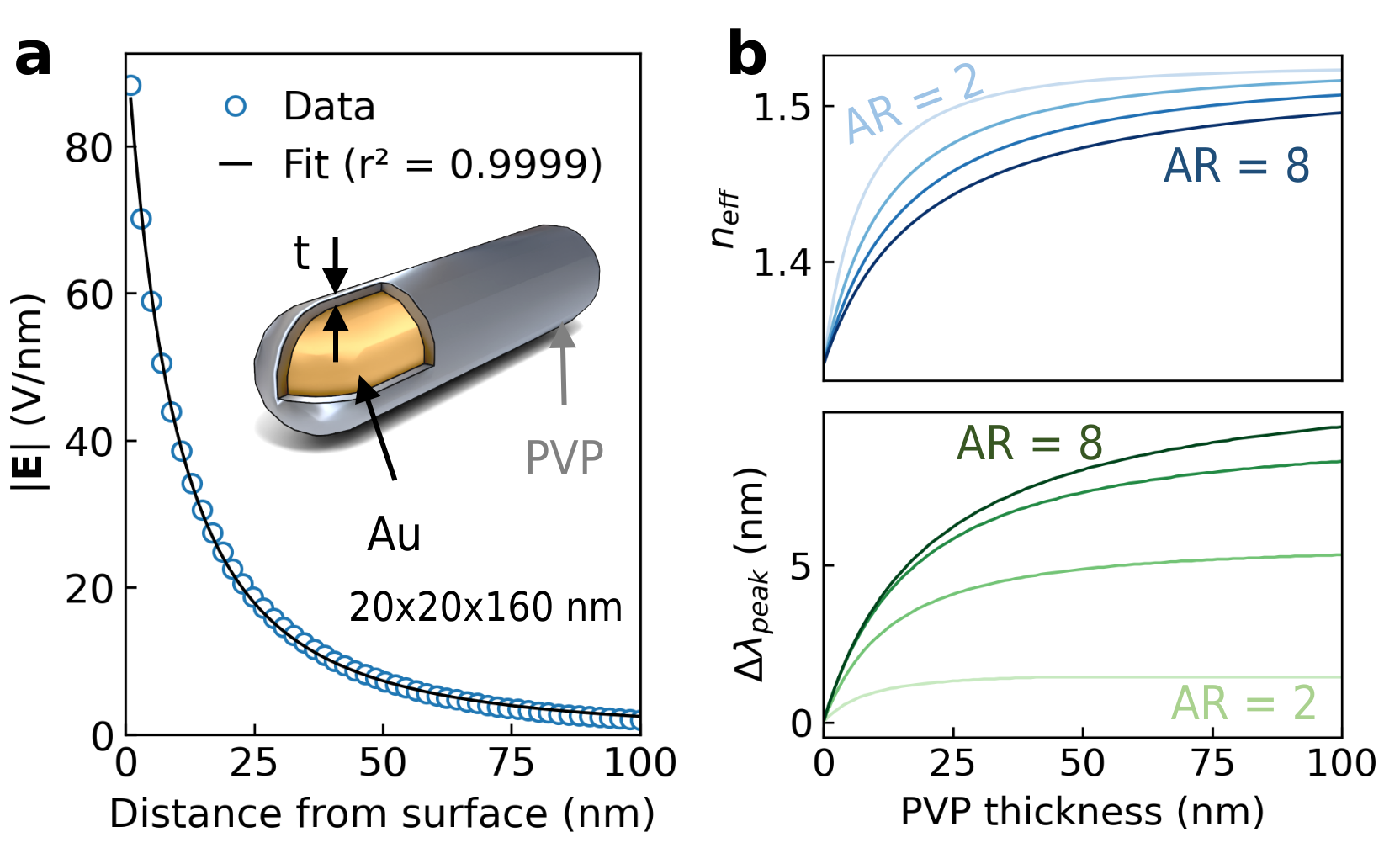}
\caption{
\(\mathbf{a}\) Integrated spatial decay profile of the electric field magnitude $|\mathbf{E}|$ with respect to distance to the surface for a gold nanorod ($20\times20\times160$~nm), extracted from numerical simulations and fitted with the power-law decay [Eq.~(\ref{eq. fit})].
\(\mathbf{b}\) Top: effective refractive index $n_{\mathrm{eff}}$ sensed by the gold nanorods with aspect ratios between 2 and 8, at constant width of 20 nm, calculated using Eq.~(\ref{eq:neff_powerlaw}). Bottom: resulting LSPR wavelength shift $\Delta\lambda$ as a function of aspect ratio, for several PVP coating thicknesses.
}
\label{fig:bio}
\end{figure}

\begin{figure}[!ht]
\centering
\includegraphics[width=0.55\columnwidth]{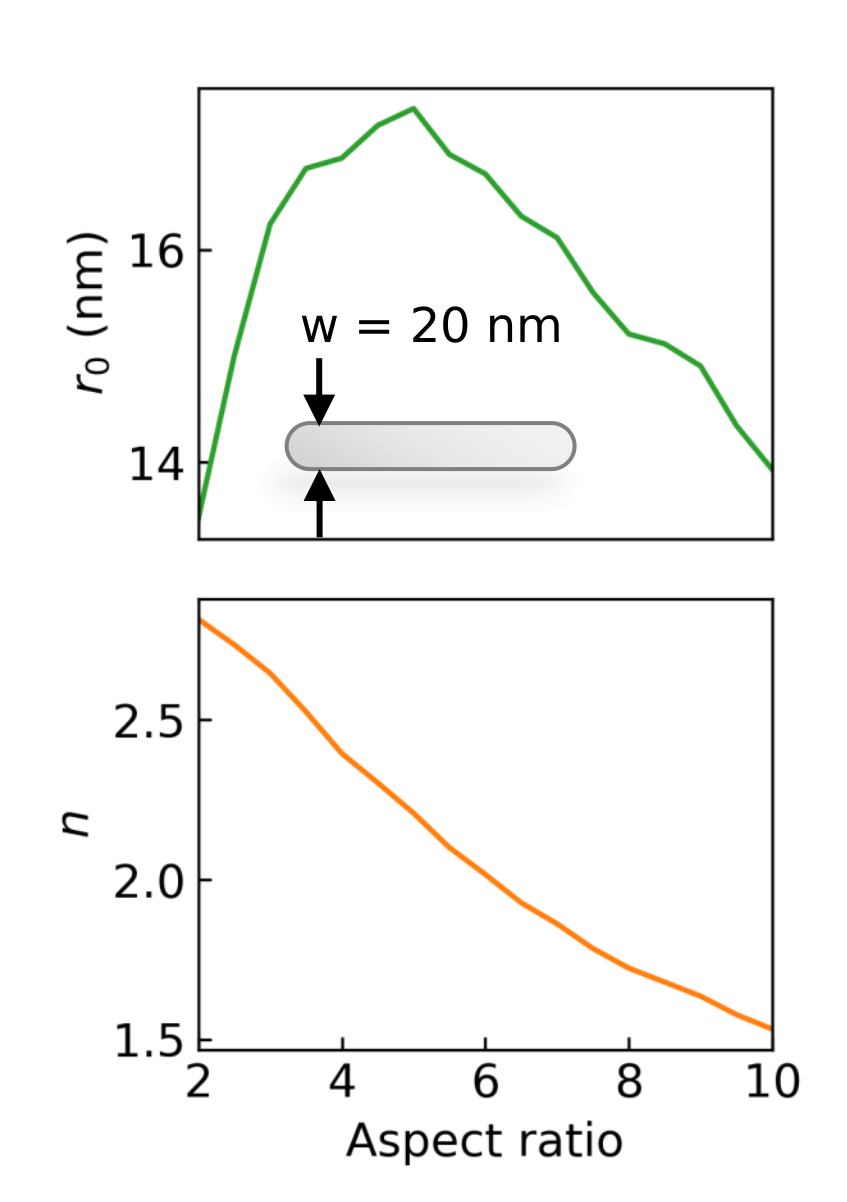}
\caption{Evolution of the fitted parameters $r_0$ and $n$ for aspect ratios between 2 and 10 for nanorods of the same width (20 nm).}
\label{SI: fig n_vs_ar}
\end{figure}

As a representative example, consider a single gold nanorod ($20 \times 20 \times 160$) aaligned along the z-axis, immersed in water, and uniformly coated with a homogeneous PVP layer of thickness \(t\) and refractive index 1.527, with the surrounding medium remaining water. First the field decay profile $|\mathbf{E}|$ is evaluated by integrating the field at constant distances and fitting the data to Eq.~(\ref{eq:neff_powerlaw}). Results displayed in Fig.~\ref{fig:bio} \(\mathbf{a}\) show the following fitted parameters: {$r_0$ = 19.0, n = 1.99}. Here, $r_0$ broadly follows the width of the nanorod, and \(n\) steadly decreases with aspect ratio increase (See S.I. Fig.~\ref{SI: fig n_vs_ar}). At low aspect ratios, the system behaves similar to two point dipoles in proximity where the field decay is proportional to $d^3$, as the nanorod spherical caps are close to each other. Inversely, at larger aspect ratios, the spherical caps are well separated, and consequently $n$ decreases. In this case, $n \approx 1.5$ at an aspect ratio of 10 was observed.
  
The dependence of the decay exponent $n$ on the nanorod aspect ratio reveals an important geometric trade-off. Increasing the aspect ratio enhances the longitudinal dipolar eigenvalue and thus the refractive-index sensitivity, but simultaneously leads to a slower spatial decay of the near field, extending its penetration depth into the surrounding medium. A consequence, of the $n$ inverse proportionality on aspect ratio is the decrease on $n_{eff}$ with respect to coating thickness, as displayed on the top panel in Fig.~\ref{fig:bio}\(\mathbf{b}\). This competing behavior results in a non-monotonic dependence of the net wavelength shift $\Delta\lambda$ on aspect ratio, as shown in the lower panel of Fig.~\ref{fig:bio}\(\mathbf{b}\).

These results demonstrate that refractive-index sensitivity alone is insufficient to predict sensing performance. Instead, optimal sensor design requires balancing geometric enhancement of the dipolar eigenmode with the spatial extent of the near field relative to the thickness and refractive index of the coating layers. The power-law effective-medium framework introduced here provides a compact and physically transparent route to achieve this optimization, enabling predictive design of plasmonic biosensors across a broad range of nanoparticle geometries.

\subsection{Computational performance}
\label{subsec:performance}

\begin{figure}[!ht]
\centering
\includegraphics[width=0.85\columnwidth]{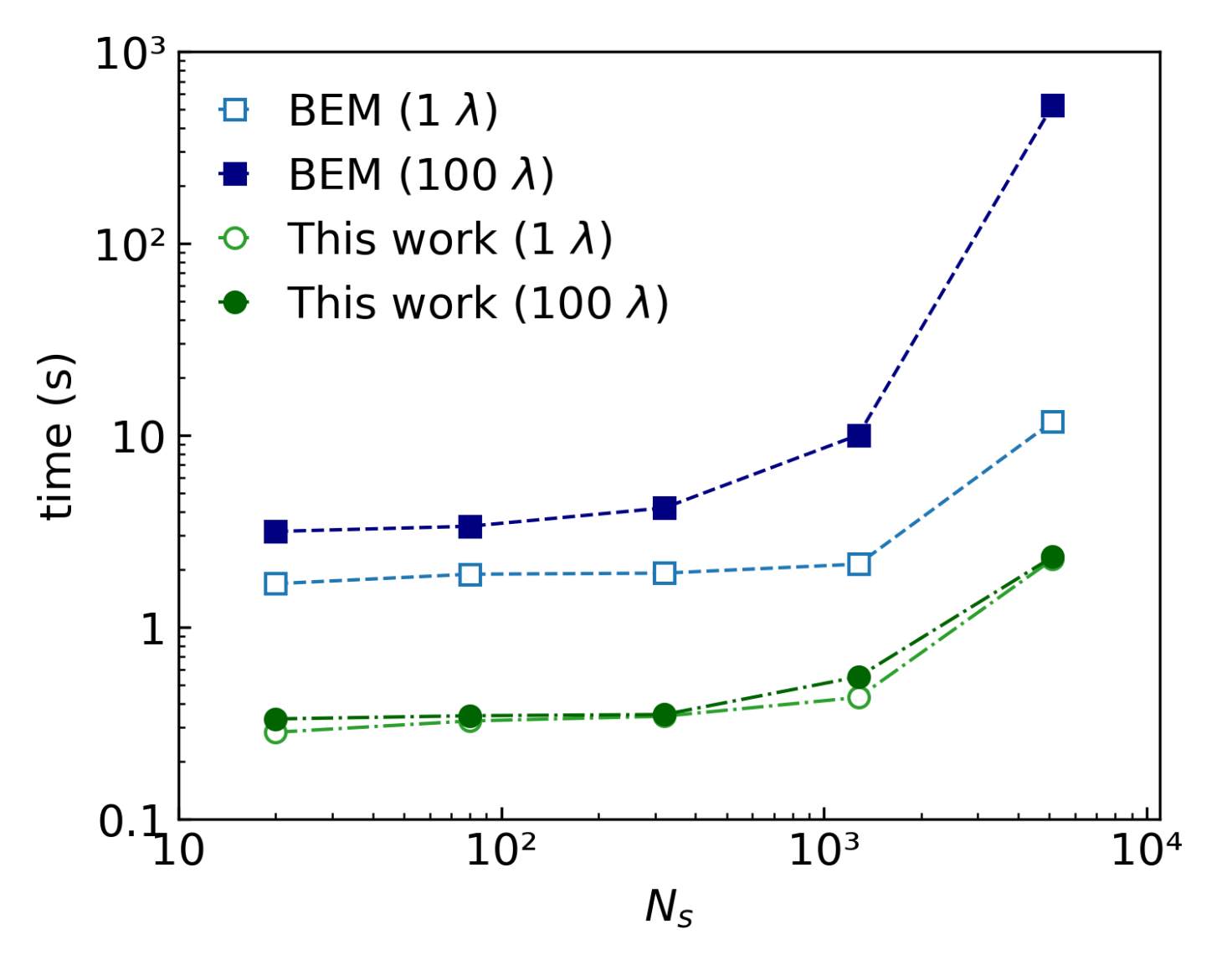}
\caption{
Computational performance comparison between a conventional boundary-element method (BEM) in the quasistatic regime and the present approach.
Wall-clock runtime is shown as a function of the number of surface elements $N_s$ for single-wavelength (1$\lambda$) and multi-wavelength (100$\lambda$) spectral evaluations. For BEM, the system matrix is assembled and solved independently at each wavelength, leading to a strong increase in runtime for spectral sweeps.
In contrast, the present method separates geometry-dependent and material-dependent contributions: geometry-dependent quantities are computed once, while subsequent spectral evaluations require only scalar operations. As a result, the runtime of the present method exhibits only a weak dependence on the number of evaluated wavelengths, enabling orders-of-magnitude speedups for large meshes and multi-wavelength sweeps.
}
\label{fig:performance}
\end{figure}

We assess the computational efficiency of the proposed method by comparing its runtime against the BEM implementation by MNPBEM~\cite{Hohenester2012} in the quasistatic approximation, for tasks relevant to plasmonic spectral analysis and sensor optimization. The goal of this comparison is not to replace full-wave solvers, but to quantify the computational advantage obtained when geometry-dependent quantities are precomputed and reused for repeated spectral evaluations. Figure~\ref{fig:performance} shows the wall-clock runtime as a function of the number of surface elements $N_s$ for both methods, considering single-wavelength (1$\lambda$) calculations and spectral sweeps over 100 wavelengths (100$\lambda$). All benchmarks were performed on the same hardware and using the same nanoparticle meshes. For BEM, the system matrix must be assembled and solved independently at each wavelength, resulting in a runtime that increases both with mesh resolution and with the number of evaluated wavelengths. This behavior is evident from the pronounced separation between the 1$\lambda$ and 100$\lambda$ curves in Fig.~\ref{fig:performance}, particularly for large $N_s$.

In contrast, as the present approach explicitly separates geometry-dependent and material-dependent contributions, all geometry-dependent quantities, including the interaction operator and its dipolar projections, are computed once for a given mesh during a preprocessing step. Subsequent spectral evaluations involve only scalar operations that depend on the material permittivity, leading to a runtime that is nearly independent of the number of wavelengths considered. As a result, virtually no computing time difference between the 1$\lambda$ and 100$\lambda$ is observed across the entire range of $N_s$.

For large meshes and multi-wavelength sweeps, and changing the surrounding medium, as done in the previous section, this separation leads to substantial computational savings. At the largest mesh sizes considered, speedups exceeding one to two orders of magnitude are observed relative to quasistatic BEM for 100-wavelength spectral evaluations. Such performance gains are particularly relevant for parameter sweeps and geometry optimization tasks, where the resonance wavelength or refractive-index sensitivity must be evaluated repeatedly.

While we emphasize that full eigensolvers in the retarded regime remain indispensable for full-wave simulations, the present method allow for unprecedented fast parametric exploration. In this context, the term \emph{ultrafast} refers to the near-instantaneous spectral evaluation enabled by one-time geometry preprocessing, making the approach especially well suited for iterative design and optimization of plasmonic nanostructures.

\section{Discussion}
\label{sec:discussion}

The results presented in this work establish a unified and computationally efficient framework for modeling the dominant dipolar plasmonic response of metallic nanoparticles with arbitrary geometry. By combining a dipole-subspace projection of the Neumann--Poincar\'e surface operator with a geometric reconstruction of the induced surface charge, the method achieves a clear separation between geometry-dependent and material-dependent contributions to the optical response. This separation underlies both the physical transparency of the approach and its exceptional computational performance.

A central outcome of the present formulation is the identification of the dipolar K eigenvalues $\kappa_n$ as intrinsic, geometry-only descriptors that govern the resonance condition, refractive-index sensitivity, and modal anisotropy of plasmonic nanoparticles. Unlike effective depolarization factors derived from geometric moments, which provide only an approximate description of resonance positions, the projected K eigenvalues recover the exact analytical spectrum for ellipsoids and extend naturally to realistic, non-ellipsoidal shapes such as nanorods, nanodisks, nanobipyramids, and nanorings. This demonstrates that the dipole-subspace K projection captures the essential electrostatic physics of dipolar plasmon modes while avoiding the computational burden of full boundary-integral eigenvalue problems.

The introduction of the geometric amplification factor $g(\kappa)=(1+2\kappa)/(1-2\kappa)$ provides a compact and physically intuitive link between nanoparticle geometry and refractive-index sensitivity. The nonlinear divergence of $g(\kappa)$ as $\kappa \to 1/2$ explains the strong sensitivity enhancement observed for elongated or sharp-featured nanoparticles, while simultaneously clarifying why transverse modes exhibit weak and often saturating sensitivity with increasing aspect ratio. Importantly, the explicit expression for the refractive-index sensitivity derived in Eq.~(\ref{eq:ris_explicit}) shows that the commonly observed linear dependence of the LSPR wavelength on refractive index arises naturally as a first-order perturbative response, despite the presence of an explicit $n_d$ factor. Genuine nonlinear sensing behavior emerges only when geometric eigenvalues become environment-dependent or when strong material dispersion curvature is present.

Beyond spectral response, the ability to reconstruct dipolar surface charge distributions directly from geometry enables efficient and self-consistent near-field mapping. The comparison with full BEM solutions confirms that the geometric-moment reconstruction faithfully reproduces the global dipolar charge pattern and associated near-field localization, with discrepancies confined to regions where higher-order multipoles become significant. This behavior is expected and highlights a key strength of the method: it provides an accurate description of the dominant plasmonic physics while explicitly identifying the regime where multipolar effects limit the dipole approximation.

The power-law description of the near-field decay further extends the utility of the framework to biosensing applications. The proposed weighting scheme captures how both geometric enhancement and near-field penetration depth jointly determine sensor performance, where optimal designs must balance eigenvalue-driven sensitivity enhancement against the spatial overlap between the plasmonic near field and the finite-thickness coating layers. The present framework provides a predictive and physically transparent route to identify this balance without resorting to repeated full-wave simulations.

From a computational standpoint, the separation between geometry preprocessing and spectral evaluation enables orders-of-magnitude speedups relative to conventional BEM approaches for multi-wavelength sweeps and optimization tasks. While BEM remains indispensable for fully retarded simulations, detailed multipolar analysis, and strongly coupled systems, the present method is ideally suited for rapid exploration of parameter space, inverse design, and high-throughput optimization of plasmonic nanostructures operating in the dipole-dominated regime. In this context, the term \emph{ultrafast} refers not to single-frequency performance, but to the near-instantaneous evaluation of spectra once geometry-dependent quantities have been computed.

Several limitations and extensions of the present approach merit discussion. First, the accuracy of the method relies on the dominance of dipolar modes; for nanoparticles with dimensions comparable to the wavelength or for strongly multipolar resonances, higher-order modes must be included. Second, the MLWA captures only first-order retardation effects and cannot fully reproduce mode hybridization or radiation-induced spectral reshaping in the deeply retarded regime. Third, although the current formulation assumes a homogeneous embedding medium, the framework can eventually be systematically extended to include planar substrates and layered environments through image-charge constructions and modified Green's functions.

Overall, the proposed dipolar electrostatic framework bridges the gap between fully numerical solvers and overly restrictive analytical models. By retaining geometric generality, physical interpretability, and exceptional computational efficiency, it provides a powerful tool for the rational design and optimization of plasmonic nanoparticles, particularly in sensing applications where repeated spectral evaluations and rapid geometry iteration are essential.

The present framework is designed to accurately capture the dominant dipolar response of subwavelength plasmonic nanoparticles operating in the weakly retarded regime. Its validity relies on the assumption that the optical response is governed primarily by dipolar modes, with higher-order multipoles contributing only weakly to far-field observables and near-field distributions. For nanoparticles with dimensions approaching or exceeding the wavelength, or for geometries that support strong multipolar resonances, the dipole-subspace approximation and the MLWA correction become insufficient to fully describe mode hybridization and retardation effects. In such regimes, full-wave electrodynamic methods remain necessary. Additionally, the current formulation assumes a homogeneous embedding medium; extensions to supported or layered environments require modified Green's functions or image-charge constructions, which can be incorporated within the same projected-operator framework. Within these well-defined limits, the method provides a physically transparent and computationally efficient tool for geometry-driven plasmonic modeling and optimization.

\section{Conclusions}

We have presented an ultrafast electrostatic framework for modeling the dominant dipolar plasmonic response of metallic nanoparticles with arbitrary geometry. By projecting the Neumann--Poincar\'e surface operator onto a Cartesian dipole subspace, geometry-dependent eigenvalues $\kappa_n$ are extracted without solving large boundary-integral eigenproblems, enabling closed-form evaluation of polarizability, near fields, and extinction spectra. The introduction of the geometric amplification factor $g(\kappa)$ provides a transparent link between nanoparticle shape and refractive-index sensitivity, clarifying why elongated and sharp-featured geometries exhibit enhanced sensing performance. 

The separation between geometry preprocessing and spectral evaluation results in orders-of-magnitude speedups for multi-wavelength sweeps and optimization tasks compared to conventional boundary-element methods, while retaining near-BEM accuracy for the dominant LSPR peak in the weakly retarded regime. By combining this spectral framework with a power-law description of the near-field decay, we demonstrate a predictive strategy for biosensor optimization that balances intrinsic sensitivity enhancement with effective near-field overlap. Together, these results establish the proposed approach as a powerful and efficient tool for the rapid design and optimization of plasmonic nanostructures for LSPR-based sensing applications.

\section*{Acknowledgements}
This work is financed by Component 5 - Capitalization and Business Innovation, integrated in the Resilience Dimension of the Recovery and Resilience Plan within the scope of the Recovery and Resilience Mechanism (MRR) of the European Union (EU), framed in the Next Generation EU, for the period 2021 - 2026, within project ATE, and within the research contract CEECIND/00471/2017.

\section*{Supporting Information}

\subsection*{Derivation of refractive-index sensitivity for arbitrary nanoparticles}\label{si_ris_derivation}

The relation 

\begin{equation}
\Re\{\varepsilon_{m}(\lambda_{\mathrm{res}})\} = -\,\varepsilon_d\, g(\kappa_z)
\label{eq:res_condition_compact}
\end{equation}

separates between geometry and material contributions. This is, the nanoparticle geometry enters exclusively through $\kappa_z$, while the material response is fully contained in the dispersive function $\Re\{\varepsilon_{m}(\lambda)\}$. To make the refractive-index sensitivity explicit, we introduce the implicit function

\begin{equation}
  F(\lambda,n_d) \equiv \Re\{\varepsilon_{m}(\lambda)\} + \varepsilon_d(n_d)\, g(\kappa_z) = 0,
\qquad
\varepsilon_d = n_d^2
\end{equation}

which is satisfied at the resonance wavelength $\lambda=\lambda_{\mathrm{res}}$. Here, $F(\lambda,n_d)$ is an implicit resonance function whose zeros define the LSPR condition. Therefore, the resonance wavelength $\lambda_{\mathrm{res}}$ is implicitly determined by the constraint $F(\lambda_{\mathrm{res}},n_d)=0$. Implicit differentiation of $F(\lambda_{\mathrm{res}},n_d)=0$ with respect to $n_d$ yields

\begin{equation}
\frac{\partial F}{\partial \lambda} \frac{d\lambda_{\mathrm{res}}}{dn_d} + \frac{\partial F}{\partial n_d} = 0
\end{equation}

and therefore

\begin{equation}
\frac{d\lambda_{\mathrm{res}}}{dn_d} = - \frac{\dfrac{\partial}{\partial n_d} \!\left(n_d^2 g(\kappa_z)\right)} 
{\left. \dfrac{d}{d\lambda} \Re\{\varepsilon_{m}(\lambda)\} \right|_{\lambda=\lambda_{\mathrm{res}}}}
\label{eq:ris_general}
\end{equation}

In the quasistatic limit for homogeneous environments, the dipolar eigenvalue $\kappa_z$ is purely geometric and independent of the surrounding dielectric medium. Consequently, $g(\kappa_z)$ is constant with respect to $n_d$, and Eq.~(\ref{eq:ris_general}) reduces to

\begin{equation}
\frac{d\lambda_{\mathrm{res}}}{dn_d} = -\frac{2 n_d\, g(\kappa_z)}{\left.\dfrac{d}{d\lambda}\Re\{\varepsilon_{m}(\lambda)\}\right|_{\lambda=\lambda_{\mathrm{res}}}}.
\label{eq:ris_explicit}
\end{equation}

\normalsize
\bibliography{references}

\clearpage
\section*{Supplementary Figures}

\end{document}